\documentclass[a4paper, 12pt]{article}
\usepackage{fullpage} 
\usepackage{setspace}
\usepackage{amsmath}
\usepackage{graphicx}
\usepackage{dsfont}
\usepackage{upgreek}
\usepackage{mathtools}
\usepackage{enumitem}

\usepackage[justification=justified,
            format=plain]{caption}
\usepackage[numbers,sort&compress]{natbib}
\allowdisplaybreaks

\usepackage{color}
\usepackage{xcolor}

\usepackage{tabularx}
\usepackage{longtable}

\usepackage{floatrow}
\usepackage{titlesec}
\titleformat*{\section}{\large\bfseries}

\usepackage{lineno}

\renewcommand*{\eqref}[1]{equation~\ref{#1}}

\title{Glia}
\author{
        Maurizio De Pitt\`a \\
            The University of Chicago, USA\\
            EPI BEAGLE, INRIA Rh\^{o}ne-Alpes, France       
        }
\date{\today}

\begin{document}
\maketitle
\vspace{1cm}
\begin{minipage}{\textwidth}
	\centering
	(Submitted as contributed section to the chapter on ``Neurophysiology'' of the book ``\textit{Da\~no cerebral}'' (Brain damage), JC Arango-Asprilla \& L Olabarrieta-Landa eds., Manual Moderno Editions.)  
\end{minipage}

\newpage
\section*{Glial cells are vital for survival and function of neural circuits}
Glial cells occupy more than a half of the volume of the human brain, and are of several types in the central nervous system (CNS), including astrocytes, oligodendrocytes and microglia \citep{Barres_Neuron2008}. Astrocytes and oligodendrocytes lineage cells are derived from neural stem cells, whereas microglia originate from the immune system \citep{VolterraMeldolesiRev2005,Hanisch_NN2007}. Moreover, in the peripheral nervous system, there are two classes of Schwann cell (myelinating and non-myelinating) which functionally and antigenically resemble the glia of the CNS \citep{Feng_COP2007}.

Glia are vital for the survival and function of neurons. Oligodendrocytes and Schwann cells myelinate axons to ensure fast, saltatory conduction of action potentials. Astrocytes regulate blood flow, provide much-needed energy to neurons, and supply building blocks of neurotransmitters, which fuel synapse function \citep{Barres_Neuron2008}. Yet the functions of glia are not restricted to supporting neurons \citep{VolterraMeldolesiRev2005}.

In the peripheral nervous system, synapses are ensheathed by non-myelinating Schwann cells, and in the CNS by astrocytes. The CNS also contains two forms of elongated, radial glial cell: Bergmann glia in the cerebellum and M\"uller cells in the retina. These have many features in common with astrocytes and are closely associated with synapses \citep{Rreichenbach_BRR2010}. This structural association extends to function. Perisynaptic glia ensure potassium ion homeostasis and regulate extracellular pH. Moreover these cells express several receptors for neurotransmitters, enabling them to `listen' to synapse function and respond to synaptic activity by making localized and global changes in intracellular calcium ion concentrations \citep{VolterraMeldolesiRev2005,Bindocci_Science2017}. In addition, glia modulate the properties of synapses by releasing neurologically active substances including hormones and neurotransmitters like glutamate, GABA, ATP and \textsc{d}-serine \citep{Araque_Neuron2014}. The extensive structural and functional association of perisynaptic glia with the synapse gave rise to the concept of the `tripartite synapse', in which synapses are defined as comprising the presynaptic and postsynaptic specializations of the neurons and the glial process that ensheaths them \citep{AraqueHaydonTripartite1999}.

\section*{Astrocytes are intimately associated with synapse formation, function and deletion}
Astrocytes are the main population of glia in the brain and have traditionally been classified into two groups on the basis of their morphology and location. \textit{Protoplasmic astrocytes} are found throughout all grey matter and exhibit several primary branches that give rise to secondary finely branching processes that ensheath synapses. \textit{Fibrous astrocytes}, on the contrary, are found in the white matter and exhibit a morphology of many long fiber-like processes that are in contact with the nodes of Ranvier \citep{KettenmannRansom_Book2013}.

Both protoplasmic and fibrous astrocytes, may form networks connecting by aqueous channels termed \textit{gap junctions}, which are site of direct intercellular communication. In this fashion, chemical signals as well as ions like potassium, sodium, chloride and calcium may travel through astrocyte networks far apart from their source site \citep{Giaume_etal_NatureRevNeurosci2010}. Remarkably, cells in these networks tile the neuropile in a contiguous and essentially non-overlapping manner that is orderly and well organized in the healthy CNS. On the contrary, astrocytic networks with a different cell arrangement or altered gap junction connections hallmark the onset of pathology \citep{Sofroniew_AN2010}.

Although the developmental generation of astrocytes tends to occur after the initial production of neurons in many CNS regions, astrocytes exert a number of important functions during development of both gray and white matter. Molecular boundaries formed by astrocytes take part indeed in guiding the migration of developing axons and certain neuroblasts \citep{Powell_Glia1999}. In addition, substantive evidence is accumulating that astrocytes are essential for the formation, function and deletion (or pruning) of synapses \citep{ErogluBarres_Nature2010}. In this fashion, astrocytes can powerfully influence synaptic remodeling and pruning at the basis of function, or lack thereof, of neural circuits in the healthy and diseased CNS \citep{Barres_Neuron2008}. Moreover, the perisynaptic processes of astrocytes can act as physical barriers for diffusion into the extracellular space of synaptically-released neurotransmitters in favor of a specificity of synaptic transmission. Perisynaptic astrocytic processes are indeed enriched in transporters that assure rapid and efficient removal of synaptic neurotransmitters, in particular glutamate and GABA. The control of the speed and the extent of neurotransmitter clearance by astrocytes could also have a role in the plasticity of synapses, because it could set the degree of postsynaptic receptor activation and desensitization \citep{BerglesRothstein2004}. On the contrary, in the presence of glutamate toxicity whereby nerve cells are damaged or killed by excessive glutamatergic stimulation -- a dangerous process which can occur in many pathological conditions including spinal cord injury or traumatic brain injury, stroke and common neurodegenerative diseases -- astrocytic glutamate transporters can quickly sequester excess glutamate, thereby providing a neuroprotective action against neuronal damage \citep{McNaught_BP2000}.

\enlargethispage{\baselineskip}
\section*{Glia-vascular coupling}
Astrocytes make extensive contacts and have multiple bidirectional interactions with blood vessels by specialized processes known as endfeet. In this fashion, astrocytes can function as relay cells in neurovascular communication, insofar as they surround synapses and are stimulated by neural activity, whereas their endfoot processes envelop blood vessels and can signal to the smooth muscle cells that control vessel diameters \citep{Attwell_Nature2010}.

Astrocyte indeed produce and release various molecular mediators, most notably arachidonic acid and its metabolites that include prostaglandins (PGEs) and epoxyeicosatrienoic acids (EETs), that can increase or decrease CNS blood vessel diameter and blood flow in a coordinated manner \citep{Gordon_Glia2007,Iadecola_NN2007}. Importantly, these mediators can be produced in response to glutamate released from synapses, and titrate blood flow in relation to levels of synaptic activity, entering blood flow by astrocyte endfoot processes. This makes astrocytes major components, along with neurons themselves, in mechanisms of functional hyperaemia, namely those mechanisms deployed by the brain to increase the flow of blood to regions where neurons are active \citep{Attwell_Nature2010}.

Astrocytes can also take up glucose from blood by their endfoot processes, thereby providing energy metabolites to different neural elements in gray and white matter. Astrocytes are in fact the principal storage sites of glycogen granules in the CNS which can be used to sustain neuronal activity during hypoglycemia and during periods of high neuronal activity \citep{Brown_Glia2007,Suh_JPET2007}. Moreover, during hypoglycemia, astrocytic glycogen can also break down to lactate which is transferred to adjacent neuronal elements -- synapses in gray matter and axons in white matter -- where it fuels aerobic metabolism therein \citep{Magistretti_ExpPhysiol2011}.

Astrocytic endfeet, along with pericytes and endothelial cells, ultimately form the blood-brain barrier (BBB) which represents the interface between the central nervous capillaries and the extracellular fluid of neurons and glial cells \citep{Barres_Neuron2008}. Although the role of astrocytes in the BBB remains controversial at the current state of knowledge, several evidence \textit{in vitro} suggest that they have a crucial role both in the induction and in the maintenance of the BBB. Several studies \textit{in vitro} indeed suggest that astrocytes could induce BBB epithelial cell types \citep{Haseloff_CMN2005,Nag_Book2011}. It should be noted however, that this induction is probably dependent on the functional status of astrocytes. Experiments in mice revealed in fact that a compromised BBB correlates with altered astrocytes that are unable to induce BBB properties \citep{Liedtke_Neuron1996,Pekny_Glia1998}.

\section*{Reactive astrogliosis and glial scar formation}
Given the ubiquity of glia cells in the CNS, almost any CNS pathology could be regarded, to some extent, as a pathology of glia \citep{Barres_Neuron2008}. Molecular, cellular and functional changes in astrocytes that occur in response to all forms and severities of CNS injury and disease, including small perturbations, are collectively known as \textit{reactive astrogliosis}. While the different severities of reactive astrogliosis seamlessly transition along a continuum, it is convenient, for the sake of description and classification, to distinguish between mild and moderate reactive astrogliosis versus severe reactive astrogliosis \citep{Sofroniew_AN2010}.

\textit{Mild to moderate reactive astrogliosis} correlates with variable degrees of hypertrophy of cell body and stem processes, without astrocyte proliferation and loss of individual astrocyte domains \citep{Sofroniew_TiNS2009,Sofroniew_AN2010}. This generally is the case for mild nonpenetrating and noncontusive trauma, or diffuse innate immune activation caused by viral or systemic bacterial infections, as well as for areas that are at some distance from focal CNS lesions. If the triggering mechanism is able to resolve, then mild or moderate reactive astrogliosis can also  spontaneously resolve, with astrocytes returning to a morphological appearance similar to that in healthy tissue \citep{Sofroniew_TiNS2009}.

Severe reactive astrogliosis can either be diffusive or be associated with compact glial scar formation. \textit{Severe diffuse reactive astrogliosis} associates with pronounced hypertrophy of cell body and stem processes together with intermittent astrocyte proliferation and disruption of astrocytic network architecture by substantial overlapping of anatomical domains of individual astrocytes. This can result in long-lasting or permanent reorganization of tissue architecture with a consequent reduced potential for resolution and return to normality. This type of reactive astrogliosis is generally found in areas that extend for some distance away from sever focal lesions, or in areas responding to chronic neurodegenerative triggers, or in response to certain types of infection or seizures \citep{Sofroniew_AN2010}.

In its most extreme form, severe reactive astrogliosis associates with \textit{compact glial scarring}, that is the pronounced overlapping of astrocyte processes that interdigitate to form compact borders that surround and demarcate areas of severe tissue damage, necrosis, infection or autoimmune-triggered inflammatory infiltration \citep{Bush_Neuron1999,Faulkner_JN2004,Drogemuller_JI2008,Herrmann_JN2008,Oberheim_JN2008,Voskuhl_JN2009}. Triggering insults include penetrating trauma, severe contusive trauma, invasive infections or abscess formation, neoplasm, chronic neurodegeneraton and systemically triggered inflammatory challenges. Importantly, glial scar formation is associated with substantive tissue reorganization and structural changes that are irreversible and thus persist even when triggering insults may have resolved.
 
It must be emphasized that there is not a single mechanism for reactive astrogliosis with a simple on/off switch, but that different potential structural and functional changes associated with reactive astrogliosis are regulated separately by many different potential signaling mechanisms. For example. different signaling events have been identified that separately regulate expression of structural proteins, cell hypertrophy or cell division. Moreover, other mechanisms separately regulate expression of pro- or anti-inflammatory molecules secreted by reactive astrocytes \citep{Sofroniew_Book2013}. 

Finally, reactive astrogliosis can have both detrimental and beneficial functions in the CNS \citep{Sofroniew_AN2010}. For example, reactive scar-forming astrocytes inhibit axon regrowth, thus preventing neural regeneration, but at the same time may promote neuroprotection, blood-brain barrier repair, and the restriction of inflammation. On the other hand,  dysfunction of astrocyte and reactive astrocytes can also occur and contribute to CNS disorders, either through gain of abnormal effects or loss of normal functions. Combination of mouse models of transgenic astrocyte manipulations with experimental models of CNS injury or disease indeed showed that genetic modulations of reactive astrogliosis and scar formation can markedly alter tissue repair, disease progression and functional outcome. Ablation of astrocytes and attenuation of certain astrocyte functions exacerbate disease progression and tissue degeneration and worsen functional outcome, whereas deletion of certain astrocyte genes appears to improve outcome in some situations \citep{Sofroniew_Book2013}. Collectively such findings point toward an enormous, yet incompletely understood potential for astrocytes to contribute to, or play primary roles in disease processes, tissue repair, and functional outcome in a wide variety of clinical conditions, including stroke, epilepsy, and neurodegenerative diseases.

\section*{Microglia as an emerging immune component of astrocyte pathophysiology}
Microglia are the immune system cells of the CNS where they are ubiquitous, being evenly spaced in a networklike fashion throughout the brain and spinal cord, although, unlike astrocytes, they are not known to normally form connections with each other \citep{Streit_Book2013}. Functionally speaking, one might think of microglia as hybrid cells that work as ``sensors of pathology'' in the brain and combine characteristics of neuroprotective glial cells with some of the attribute of white blood cells like macrophages and lymphocytes. Indeed like astrocytes, microglia can exist either in the resting (ramified) state, or in the reactive (hypertrophic) one, and like white blood cells, can also be phagocytic \citep{Streit_Glia1988}.

A well known case of phagocytic microglial is observed in animal models of transection of the facial nerve which results in retrogade injury of hypoglossal or spinal motorneuron cell bodies. After axotomy, reactive microglia adhere to damaged motorneurons, actively removing neuron-neuron (nerve) connections in a way that breaks down damaged synapses \citep{Cullheim_BRR2007}. On the other hand, reactive microgliosis is not necessarily phagocytic and destructive. A common hallmark of microgliosis that is microglia migration to and proliferation at the injury site, is rather neuroprotective. Similarly to astrocytes in fact, microglia express high levels of glutamate transporters in the area surrounding the injured motorneuron, protecting this latter from the abnormal glutamate-excitation that would be otherwise induced \citep{Lopez_MBR2000}.

Very often reactive microgliosis occurs together with reactive astrogliosis \citep{Cullheim_BRR2007}. The mechanism underpinning such correlated reactivity of astrocytes and microglia is a matter of active investigation, yet it seems to be caused by the existence of an activation loop in which one cell type-derived signal further stimulate the other cell type. Astrocytes have indeed been shown to promote reactive microglia proliferation \citep{Bartolini_ND2011} as well as suppress phagocytic migroglia activity \citep{DeWitt_EN1998}. Conversely, microglia could regulate astrocyte neuroprotective properties by secretion of different molecules, including cytokines that are normally observed in the presence of ongoing inflammatory responses \citep{Arnett_NN2001}.

Remarkably, this interaction between astrocytes and microglia seems not only to be limited to reactive gliosis, but also to occur in the healthy CNS. Synaptic pruning by astrocytes could be promoted by microglia signalling \citep{ErogluBarres_Nature2010}. Similarly, astrocytic regulation of synaptic glutamate release might require upstream microglia activation \citep{Pascual_PNAS2011}. Although at present these aspects have only been confirmed in experiments in rodents brain slices, they nevertheless pinpoint to the existence of potentially critical interactions between immune system cells (microglia) and astrocytes, which should be taken into account in the design of future therapeutic approaches.

\newpage
\bibliography{./glia.bib}

\end{document}